# Bayesian Hyperparameter Optimization for Deep Neural Network-Based Network Intrusion Detection

Mohammad Masum∗, Hossain Shahriar†, Hisham Haddad†, Md Jobair Hossain Faruk‡, Maria Valero†
Md Abdullah Khan§, Mohammad A. Rahman¶, Muhaiminul I. Adnan‖, Alfredo Cuzzocrea∗∗, Fan Wu††

∗School of Data Science, Kennesaw State University, USA
†Department of Information Technology, Kennesaw State University, USA
‡Department of Software Engineering and Game Development, Kennesaw State University, USA
§Department Computer Science; Kennesaw State University, USA
¶Department of Electrical and Computer Engineering, Florida International University, USA
‖Institute of Natural Sciences, United International University, USA
∗∗iDEA Lab, University of Calabria, Rende, Italy and LORIA, Nancy, France
††Department of Computer Science, Tuskegee University, USA

*Abstract*— **Traditional network intrusion detection approaches encounter feasibility and sustainability issues to combat modern, sophisticated, and unpredictable security attacks. Deep neural networks (DNN) have been successfully applied for intrusion detection problems. The optimal use of DNN-based classifiers requires careful tuning of the hyper-parameters. Manually tuning the hyperparameters is tedious, time-consuming, and computationally expensive. Hence, there is a need for an automatic technique to find optimal hyperparameters for the best use of DNN in intrusion detection. This paper proposes a novel Bayesian optimization-based framework for the automatic optimization of hyperparameters, ensuring the best DNN architecture. We evaluated the performance of the proposed framework on NSL-KDD, a benchmark dataset for network intrusion detection. The experimental results show the framework's effectiveness as the resultant DNN architecture demonstrates significantly higher intrusion detection performance than the random search optimization-based approach in terms of accuracy, precision, recall, and f1-score.**

## I. INTRODUCTION

Network Intrusion Detection Systems (NIDSs) identify computer networks' unauthorized access and investigate potential security breaches. Traditional NIDSs encounter difficulties in combating newly created sophisticated and unpredictable security attacks. While the number of security threats to traffic data increases exponentially and the attacks are becoming more sophisticated and variants. Hence, traditional intrusion detection techniques such as signature-based detection, heuristic detection, or behavior-based detection are not adequate to combat malicious activities [1].

Machine learning (ML) algorithms, including DNN, have exhibited promising results in classifying network intrusions [2]. The algorithms can overcome the limitations of traditional detection methods and provide a rewarding accuracy score.

DNN-based classifiers have gained popularity and are widely used for many applications, including computer vision, speech recognition, and natural language processing. Extracting automated features, the capability of highly non-linear systems, and flexibility in architecture design are the highlights of the DNN. However, achieving successful performance from the DNN requires careful tuning of the training- and the hyper-parameters. While the training parameters are learned during the training phase, the hyperparameters such as initial learning rate, numbers of dense layers, number of neurons, and activation functions must be specified before training the DNN. The DNN's performance significantly depends on hyperparameter optimization. It requires investigating the optimal combination of hyperparameters of the DNN model that returns the best performance as scaled on a validation set [3].

Manually tuning the hyperparameters involve expert experience. Lack of skill to manually set hyperparameters is discouraging to use DNN efficiently. Especially, when there are many hyperparameters, it is too tedious, time-consuming, and computationally expensive to model the best DNN architecture by brute-force approaches because it requires performing experiments with many possible combinations of all the hyperparameters [4]. For instance, grid search and random search are commonly used by research communities for hyperparameter optimization [5]. Grid search explores all possible combinations of hyperparameters to achieve global optimum. The random search applies several numbers of arbitrary combinations of hyperparameters. Both methods are easy to implement. However, when dealing with several hyperparameters, these methods converge slowly, take a long time, and are not guaranteed to perform as expected [4].

We may have hundreds of possible combinations of hyperparameters to be compared for obtaining optimal performance depending on the size of DNN architecture. Hence, an automatic and efficient method for hyperparameter optimization for DNN is vital. Bayesian hyperparameter optimization is a state-of-the-art automated and efficient technique that outperforms other advanced global optimization methods on several challenging optimization benchmark functions [6]. The Bayesian optimization (BO) uses surrogate models like Gaussian processes (GP) to define a distribution over objective function for approximating a true objective function. In this paper, we applied Bayesian optimization with Gaussian processes (BO-GP) for tuning hyperparameters of DNN. In summary, the contribution of this analysis is two-fold:

- We proposed a novel network intrusion detection framework by optimizing DNN architecture's hyperparameters leveraging Bayesian optimization.

- We evaluated the proposed framework and compare it with the random search hyperparameter optimization method using a benchmark network intrusion dataset.

The rest of the paper is organized as follows: In Section II, we discuss ML-based related work of network intrusion detection. Section III explains the methods we applied in this paper. The experimental setting and results are explained in Section IV. Finally, Section V concludes the paper.

## II. RELATED WORK

First, Addressing the constraints of traditional methods, researchers have proposed both conventional machine learning (ML) algorithms and deep learning for network intrusion detection. Deep neural networks are successfully applied for classifying malware activities and network intrusion detection. A two-stage deep neural network approach was performed for intrusion detection of the KDD CUP'99 dataset, where an artificial neural network was applied for feature selection in the initial stage and an autoencoder was applied for classification in the final part [7]. A multilevel classifier fusion approach was proposed for network intrusion detection, containing two layers wherein the upper layer, an unsupervised feature extraction method was applied using a non-symmetric deep autoencoder, and a random forest classifier was then applied with the extracted features [8]. The method experimented with two different datasets including the KDDCUP'99 and the NSL-KDD datasets. For a 5-class classification, the proposed method achieved 85.42% accuracy for the NSL-KDD dataset.

Limited numbers of published works are available that are based on hyperparameter optimization for network intrusion detection. Different hyperparameter optimization techniques including random search, meta-heuristic algorithms, and Bayesian optimization are applied to optimize hyperparameters for traditional ML methods like K-Nearest Neighbor (KNN), and Random Forest (RF) classifiers for network intrusion detection [9]. The authors optimized- the number of neighbors for KNN and splitting criteria, and the number of trees for RF. Experimental results in this study showed significant improvement in detection accuracies with hyperparameter optimization. BO technique was implemented for optimizing hyperparameters of support vector machine, RF, and K-NN classifiers towards intrusion detection [10]. A decision tree classifier optimized with BO-GP was applied to detect botnet attacks [11]. Other optimization techniques like particle swarm optimization (PSO), weighted local search (WLS), and genetic algorithms (GA) were applied for improving the performance of intrusion detections. A PSO-based optimization technique was applied for tuning 12 different hyperparameters (both global and local) to search for the best neural network architecture for intrusion detection [12]. In this paper, we applied the BO-GP hyperparameter optimization technique to examine the best DNN architecture for network intrusion detection".

## III. METHODS

We applied a DNN architecture with hyperparameter optimization for network intrusion detection. In the first section, the DNN method is explained, and in the next section, we explain the hyperparameter optimization techniques that are applied in this paper. In the final section, we discuss the proposed framework that integrates DNN architecture with Bayesian Optimization.

### A. Deep Neural Network

At present, deep neural networks are widely used for many applications due to the capability of highly non-linear systems and flexibility in architecture design. The neural network's basic architecture contains input layers, one or multiple hidden layers, and output layers where each of the layers includes a certain number of neurons. Weighted linear combination of neurons of a layer is computed and then used as input to another neuron in the succeeding layer. To capture the non-linearity of the data, a non-linear function, called activation function, can be applied to the weighted sums of neurons. Rectified linear unit (ReLU), sigmoid, and hyperbolic tangent (TanH) are commonly used activation function. Equation 1, 2, and 3 describes ReLU, sigmoid, and TanH functions, respectively.

$$S(x) = \frac{e^x}{1+e^x} \qquad (1)$$

$$R(x) = \begin{cases} x, & x > 0 \\ 0, & x \leq 0 \end{cases} \qquad (2)$$

$$\text{TanH}(x) = \frac{e^x - e^{-x}}{e^x + e^{-x}} \qquad (3)$$

All the weights of a neural network are set to random values at the initial stage of training. Data is fed into the input layer of the network, then it travels through the hidden layers, and finally, the output is produced in the output layer. The network continually updates the weights applying backpropagation based on the output and desired target of the neural network. The network consequently reduces the error between the output and target in each iteration [13]. In the process, a loss function is used to calculate the error of the network, and the error is minimized by applying an optimization function during backpropagation. Gradient descent-based update function generally used for backpropagation to search optimal solution by updating weight parameters. The gradient descent learning process is expressed in equation 3, where $\eta$, $t$, and $W$ are the learning rate, number of epochs, and weight matrix, respectively. We applied binary cross-entropy as a loss function that computes intrusion detection probability between 0 and 1. Equation 4 defines the binary cross-entropy (known as log loss) where $y$, and $\hat{y}$ are ground truth and predicted value, respectively [14].

$$\arg\min H(y, \hat{y}) = -y\log\hat{y} - (1-y)\log(1-\hat{y}) \quad (4)$$

Gradient descent-based update function generally used for backpropagation to search optimal solution by updating weight parameters. The gradient descent learning process is expressed in equation 5, where $\eta$, $t$, and $W$ are the learning rate, number of epochs, and weight matrix, respectively.

$$W(t+1) = W(t) - \eta \frac{\delta H(t)}{\delta W(t)} \quad (5)$$

### B. Hyperparameter Optimization

An optimal set of hyperparameters allows performance improvement as well as avoid performance issues like overfitting.

Bayesian Optimization: Bayesian optimization (BO) is a probabilistic optimization technique that aims to globally minimize an objective black-box function $f: \mathbb{X} \rightarrow \mathbb{R}$ for some bounded set $x^* \in \arg\min_{x \in \mathbb{X}} f(x)$ [15]. The common assumption is that the black-box function $f$ has no simple closed-form but can be evaluated at any arbitrary $x \in \mathbb{X}$ [16]. Additionally, the function $f$ can only be measured by unbiased stochastic observations $y$ [15]. The hyperparameter design space of interest, $\mathbb{X}$, can include continuous, integer-values, or categorical. In the process of optimization, it constructs a surrogate function (probabilistic model) $p(f)$ that defines a distribution over the objective function $f$ for approximation, and an acquisition function $a: \mathbb{X} \rightarrow \mathbb{R}$ is used to quantify the effectiveness of evaluation at any $x$. In short, the BO framework consists of three key components: surrogate model, Bayesian update process, and acquisition function [7]. The surrogate model fits all the points of the objective function and is then updated by the Bayesian updated process after each new evaluation of the objective function. Finally, the acquisition function assesses the evaluation. Algorithm 1 displays the basic pseudo-code for BO [16].

The BO can include different surrogate models: GP, sequential optimization using decision trees; different acquisition functions: expected improvement (EI), lower confidence bound (LCB), and the probability of improvement (PI). In this paper, we applied GP as the surrogate model and EI as the acquisition function.

---

**Algorithm 1** Bayesian Optimization

1. Initialize data $D_0$ using an initial design
2. for $t = 1, 2, ...,$ do
    a. fit probabilistic model for f(x) on data $D_{t-1}$
    b. select $x_t$ by optimizing the acquisition function $x_t = \arg\min_{x} a_{p(f)}(x; D_t)$
    c. evaluate the objective function $y_t$
    d. augment data $D_t = \{D_{t-1}, (x_t, y_t)\}$
    e. update model $x_t^* \leftarrow \arg\min \{y_1, y_2, ..., y_t\}$
3. end for

---

Gaussian Processes: GP is a convenient and renowned selection as a probabilistic model due to the descriptive power and analytic tractability [16]. GP assumes that every finite subset of a collection of random variables follows a multivariate normal distribution, i.e., GP assumes a priory that the probability $p(f(x_1), f(x_2), ..., f(x_k))$ for a finite collection of points $x_1, x_2, ..., x_k \in \mathbb{R}^d$ follows a multivariate normal distribution with a mean function $\mu(x)$ and covariance function $\kappa(x_i, x_j)$, where $\kappa$ is a positive definite kernel such as the squared kernel, the rational quadratic kernel, and the Matern kernel [17].

Expected Improvement (EI): Acquisition function like EI investigates trade-off between exploration and exploitation and decides the subsequent evaluation point for optimization. EI is defined as: $a_{EI}(x) = E[\max(0, f(x^*) - f(x))]$, where $f(x^*)$ is the best-observed value known and $E[.]$ is the expectation specifying the function values $f(x)$. Therefore, the reward or improvement is $f(x^*) - f(x)$ if $f(x^*) > f(x)$, while the reward is zero otherwise.

## IV. EXPERIMENT & RESULTS

In this section, we present both the experiments and results. We first specify the dataset followed by data processing. In order to provide an effective experiment, we utilize accuracy, precision, recall, and F-score metrics for evaluating models' performance. We then define experimental settings using BO-GP, and random search. Lastly, we present the experimental results.

### A. Dataset Specification

NSL-KDD dataset is the refined version of the KDDCUP'99 dataset which was widely used for network intrusion detection problems. The KDDCUP'99 dataset contains concerns like redundant records that generate biases towards frequent records. In the NSL-KDD dataset, the issues were addressed. KDDTrain+, KDDTest+, and KDDTest-21are the three datasets that are included in the NSL-KDD. In this paper, we used KDDTrain+ for training and the other two datasets for testing our proposed framework. NSL-KDD dataset consists of 41 features along with a label attribute indicating the connection status (normal or attack). Table 1 displays the distribution of the three datasets for binary classification (normal vs. attack).

**Table 1: Distributions of the NSL-KDD dataset**

| Type | KDDTrain+ | KDDTest+ | KDDTest-21 |
|---|---|---|---|
| Normal | 67,345 | 97,11 | 2152 |
| Attack | 58,630 | 12,833 | 9698 |

### B. Data Preprocessing

The categorical variables in NSL-KDD mapped to numeric data at first and then the overall data was normalized. Protocol type, flag, and service are the three categorical features that are converted into numeric data using the one-hot encoder technique. We applied the min-max normalization technique to scale the original data to a fixed range of 0 and 1. The normalization ensures consistency of the data distribution and avoiding the exploding gradients problem in the training phase.

### C. Model Evaluation Metrics

Accuracy, precision, recall, and F-score metrics were used for evaluating models' performance. These metrics use properties from the confusion matrix such as true positive (TP), false positive (FP), false negative (FN), and true negative (TN). TP is the number of attacks that are correctly classified as attacks, while FN is the number of attacks that are incorrectly classified. The number of incorrectly classified normal data is FN, and TN is correctly classified as normal data. Equations 6, 7, 8, and 9 are the mathematical definition of the performance metrics accuracy, precision, recall, and F-score, respectively.

$$Accuracy = \frac{TP + TN}{TP + FP + FN + TN} \quad (6)$$

$$Precision = \frac{TP}{TP + FP} \quad (7)$$

$$Recall = \frac{TP}{TP + FN} \quad (8)$$

$$F-score = \frac{2 * Precision * recall}{Precision + recall} \quad (9)$$

### D. Experimental Setting

We applied BO-GP, and random search to explore the best DNN architecture for network intrusion detection. Table 2 displays the hyperparameters with their range values that were studied in this paper. We explored the number of dense (hidden) layers ranging from 1 to 3. In the hidden layers, we examined three hyperparameters: number of neurons in the hidden layer— between 10 and 100; dropout rate— between 0.1 and 0.6; activation function— ReLU, sigmoid, and TanH.

The input shape of the network is $121 \times 1$, which is the number of variables in the NSL-KDD data. Thus, we set the maximum number of neurons in the hidden layers to 100 as a rule of thumb— the number of hidden neurons should remain 70% to 90% of the input size [18]. Adding excessive hidden layers may lead to a complex network and cause overfitting. Hence, the maximum number of hidden layers was set to 3 to save training time and complexity [19].

**Table 2: Hyperparameters setting by exploring different ranges**

| Hyperparameter | Range |
|---|---|
| Number of dense layers | $\mathbb{Z}[1,3]$ |
| Number of neurons | $\mathbb{Z}[10, 100]$ |
| Dropout rate | $\mathbb{R}[0.1, 0.6]$ |
| Activation function | ReLU & Sigmoid |
| Optimizer | Adam & SGD |
| Learning rate | $\mathbb{R}[10^{-1}, 10^{-6}]$ |

Dropout is one of the regularization techniques that prevents overfitting and unnecessary co-adapting by randomly eliminating (dropping out) neurons during training [20]. Though a dropout rate of 0.5 is widely used, we explored different dropout rates ranging from 0.1 to 0.6. ReLU and sigmoid are widely used activation function. We included TanH along with ReLU and sigmoid functions for searching for the best activation function considering network intrusion detection. In the backpropagation part, we explored adaptive learning rate optimization (Adam), and stochastic gradient descent (SGD) for the optimizer. Finally, a learning rate ranging from 0.1 to 0.000001 was selected for optimization. We run the BO process for 40 iterations. The final layer of the DNN was the sigmoid layer as the detection problem is a binary classification.

### E. Experimental Results

We applied random search and BO-GP hyperparameters optimization techniques for DNN. Table 3 shows the optimal set hyperparameter for DNN towards network intrusion detection. The optimal hyperparameters for random search — 2 dense layers, 60 neurons in each layer, dropout rate: 0.1, activation function: ReLU, optimizer: Adam, and learning rate: 0.005. On the other hand, the achieved optimal hyperparameters for BO-GP — 1 dense layer with 100 neurons, dropout rate: 0.6, activation function: ReLU, optimizer: Adam, and learning rate: 0.0006.

Fig. 1 displays the improvement of hyperparameter optimization where the best fitness value (negated classification accuracy) is plotted on the y-axis and the number of iterations for optimization is plotted on the x-axis. Fig. 2 and 3 show the distribution of samples for activation function and optimizer during the optimization process. ReLU was sampled more than sigmoid for activation function, while Adam was sampled more than SGD for the optimizer.

Table 3: Optimal hyperparameters for random search and Bayesian optimization

| Type | Random search | BO-GP |
|---|---|---|
| Number of dense layers | 2 | 1 |
| Number of neurons | 60 | 100 |
| Dropout rate | 0.1 | 0.6 |
| Activation function | ReLU | ReLU |
| Optimizer | Adam | Adam |
| Learning rate | 0.005 | 0.0006 |

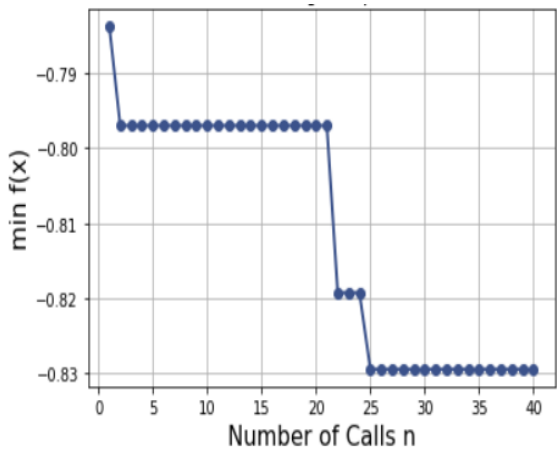

Figure 1: Convergence plot for hyperparameter optimization

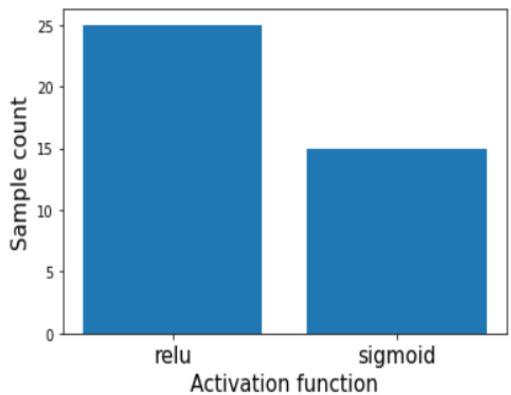

Figure 2: Distribution of samples for the activation function

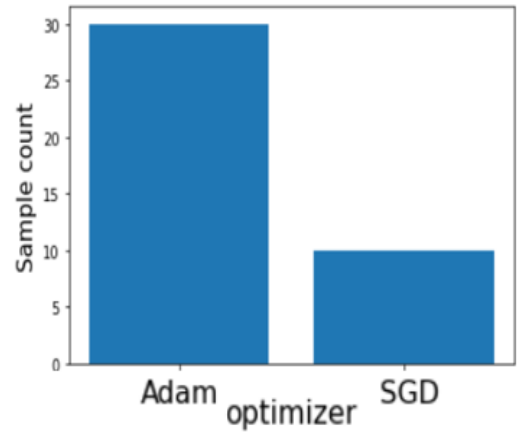

Figure 3: Distribution of samples for the optimizer

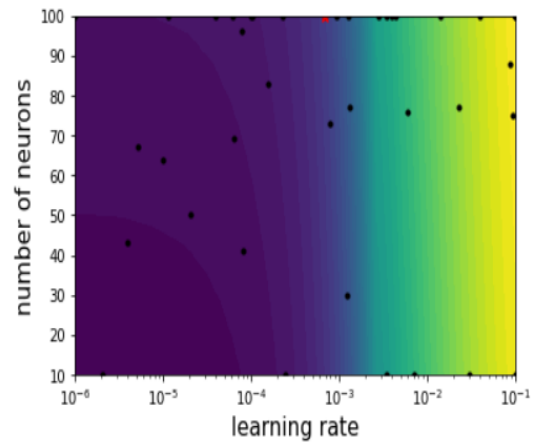

Figure 4: Landscaped plot of two optimized parameters- learning rate and number of neurons

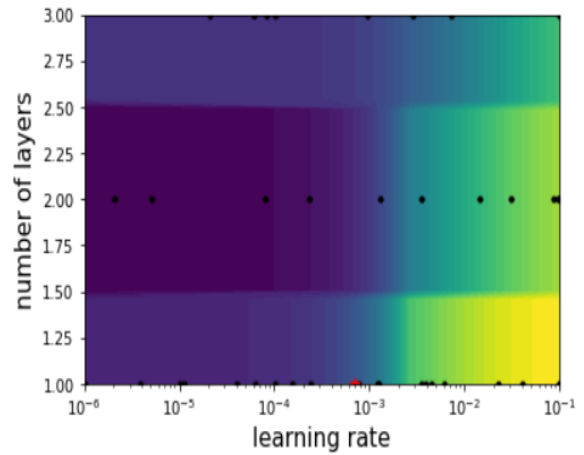

Figure 5: Landscaped plot of two optimized parameters: learning rate and number of hidden layers

Fig. 4 displays a landscaped plot of two optimized parameters: estimated fitness value for learning rate and the number of neurons in hidden layers. Black dots are the sampling location of the optimizer, while the red star is the optimal point for the two hyperparameters. Fig. 5 shows the landscaped plot for the learning rate and the number of hidden layers.

Tables 4 and 5 display the performance evaluation of KDDTest+ and KDDTest-21 datasets, respectively. The BO-GP approach achieved higher accuracy, precision, recall, and f1-score than random search optimization for KDDTest+ dataset. Moreover, the BO-GP provides better accuracy, precision, and f1-score for KDDTest-21 dataset.

Table 4: Performance evaluation of KDDTest+ data

| Type | Accuracy | Precision | Recall | F1-Score |
|---|---|---|---|---|
| Random Search | 77.46 | 72.86 | 75.99 | 74.39 |
| BO-GP | 82.95 | 79.73 | 81.35 | 80.43 |

Table 5: Performance evaluation of KDDTest-21 data

| Type | Accuracy | Precision | Recall | F1-Score |
|---|---|---|---|---|
| Random Search | 51.56 | 26.25 | 92.19 | 40.87 |
| BO-GP | 54.99 | 27.44 | 89.96 | 42.06 |

V. CONCLUSION

The efficiency of the machine learning algorithms including the deep neural network significantly depends on hyperparameter's values. Manually tuning the hyperparameters is tedious, time-consuming, and computationally expensive to model the best DNN architecture. This paper presents a framework using Bayesian optimization with Gaussian processes to optimize deep neural network hyperparameters regarding network intrusion detection. The framework was evaluated with the NSL-KDD dataset, which is a reference for the cybersecurity research community. Investigating the best DNN architecture, we explored six hyperparameters: number of hidden layers, number of neurons, dropout rate, activation function, optimizer, and learning rate. We applied a random search-based optimization technique to compare the result with our proposed approach. The experimental results showed that the BO-GP based approach outperforms the random search-based method. BO-GP achieved the highest accuracy— 82.95%, and 54.99% for KDDTest+, and KDDTest-21 datasets, respectively.

ACKNOWLEDGEMENT

The work is partially supported by the U.S. National Science Foundation Awards #2100134, #2100115, #1723578, #1723586. Any opinions, findings, and conclusions or recommendations expressed in this material are those of the authors and do not necessarily reflect the views of the National Science Foundation.